\documentclass{article}

\usepackage{amsmath}
\makeatletter % `@' now normal "letter"
\@addtoreset{equation}{section}
\makeatother  % `@' is restored as "non-letter"

\newtheorem{theorem}{\hspace{2em}Theorem}
\newtheorem{lemma}{\hspace{2em}Lemma}
\usepackage{indentfirst}
\usepackage{setspace}
\usepackage{amsmath}
\usepackage{cases}
\usepackage[colorlinks, linkcolor=black, anchorcolor=black, citecolor=black]{hyperref}
\usepackage{graphicx}
\title{Angular Momentum Independence of the Entropy Sum and Entropy Product for AdS Rotating Black Holes In All Dimensions}
\author{Hang Liu$^{a}$\thanks{E-mails:hangliu@mail.nankai.edu.cn} , Xin-he Meng$^{a,b}$\thanks{E-mails:xhm@nankai.edu.cn}
\\
\\
$^{a}$ School of Physics, Nankai University, Tianjin 300071, China\\
$^{b}$ State Key Laboratory of Theoretical Physics, Institute of Theoretical Physics,\\
Chinese Academy Of Science, Beijing 100190, China}
\date{}
\usepackage[a4paper,left=2.5cm,right=2.5cm,bottom=3cm,top=2.5cm]{geometry}
\begin{document}
\large
\maketitle
\begin{abstract}
In this paper, we investigate the angular momentum independence of the entropy sum and product for AdS rotating black holes based on the first law of thermodynamics and a mathematical lemma related to Vandermonde determinant. The advantage of this method is that the explicit forms of the spacetime metric, black hole mass and charge are not needed but the Hawking temperature and entropy formula on the horizons are necessary for static black holes, while our calculations require the expressions of metric and angular velocity formula. We find that the entropy sum is always independent of angular momentum for all dimensions and the angular momentum-independence of entropy product only holds for the dimensions $d>4$ with at least one rotation parameter $a_i=0$, while the mass-free of entropy sum and entropy product for rotating black holes only stand for higher dimensions ($d>4$) and for all dimensions, respectively. On the other hand, we find that the introduction of a negative cosmological constant does not affect the angular momentum-free of entropy sum and product but the criterion for angular momentum-independence of entropy product will be affected.
\end{abstract}
\tableofcontents
\section{Introduction}
Understanding the origin of the black hole entropy at microscopic level has been a major challenge in quantum theories of gravity (though we have not achieved the final theory over years' efforts) after the establishment of black hole thermodynamics by Hawking's radiation with black body spectrum and in analog between the black hole mechanics with the classical thermodynamics. The mass-independence of entropy sum and entropy product for a stationary black hole with multi-horizons have been studied widely in recent years, and finding the product of all horizon entropies is mass-independence in many cases, but violated  in some cases, while the more ``universal" property of mass free for general sum of all horizon entropies proposed by Meng \textit{et al.} \cite{Meng1} is preserved and only depends on the coupling constants of the theory and topology of black holes in asymptotical Kerr-Newman-(anti)-dS spacetime background, in most cases, which means that it is more general than entropy product in some cases. For Myers-Perry black holes, the entropy sum is mass dependent only in four dimensions, i.e., for the kerr solution case, which has been demonstrated in \cite{Gao1}.

For an axisymmetric and stationary Einstein-Maxwell black hole in four dimensions with angular momentum J and charge Q, Marcus Ansorg and Jorg Hennig \cite{Ansorg,Hennig} have proved the universal relation below
\begin{equation}
A_+A_- = (8\pi J)^2+(4\pi Q^2)^2
\end{equation}
where $A_+$ and $A_-$ represent the area of event horizon and Cauchy horizon respectively. Cvetic \textit{et al.} \cite{Cvetic1} have generalized the investigations by explicit calculations of the above to rotating black holes with multi-horizons in higher dimensions. They have demonstrated that the entropy product of all horizons for rotating multi-charge black holes in four and higher dimensions is mass free in either asymptotically flat or asymptotically AdS spacetime. Entropy sum of all black horizons in similar cases is also investigated in \cite{Meng1,Tian1}, which has shown that it is also independent of mass.

So far, most of the research works are focused on the mass independent properties of entropy sum and product of all horizons for kinds of black holes in variety of theories of gravity, including modified gravity theories \cite{Meng3}, as we have briefly  introduced above. Below we will present the key motivation why we keep on studying the mathematical physics related to mass and angular momentum-free relations for black holes.
Considering that a massive star with mass $ M$ large enough, angular momentum $J$ and charge $Q$ will collapse or two massive black holes merge into a black hole, i.e., a Kerr-Newman black hole after burning out all its  nuclear fuel and all the parameters needed to describe it then  are just the mass , angular momentum and charge mathematically, though numerous parameters and physical quantities with complicated physical processes required to describe it before or during its collapsing. This is the so- called `` no hair or three hairs theorem for black hole". If there exist some relations which are independent of angular momentum (and/or mass-free for more complete case) of a rotating star/black hole, we say these relations own universal property, which means no matter what initial conditions it starts the final state is universal (mass or/and angular momentum free). In astrophysics, physical black holes only possess mass and angular momentum as its final state characters. Motivated by this thought, besides the previously finished works on the new mass-free relations we continue studying angular momentum-independence relations by focusing on
investigating  the angular momentum-independence of entropy sum and entropy product of all horizons for rotating AdS black holes in the spacetime with a negative cosmological constant $\Lambda$ in the general dimensions $d\geq4$, when
$\Lambda\rightarrow 0$, it just goes back to Myers-Perry solutions. Based on the method developed by Gao, et al\cite{Gao1}, we find that the entropy sum is always independent of angular momentum for all dimensions and the angular momentum-free of product only holds for the dimensions $d>4$ with at least one rotation parameters $a_i=0$, while the mass-independence of entropy sum and product for rotating black holes only hold for higher dimensions ($d>4$) and for all dimensions, respectively. On the other hand, we find that the introduction of a negative cosmological constant does not affect the angular-momentum independence of entropy sum but the criterion for angular momentum-free of entropy product will be affected mathematically. From this point, the entropy sum may be more general than entropy product, just as in the case of mass-independence discussion. The clear expressions of entropy, Hawking temperature and angular velocity are required in our calculations, but the explicit forms of the mass and charge are not needed.

The present work is organized as follows. In next section we employ the first law of black hole thermodynamics and a mathematical lemma to the new black hole relations. In section three we focus the new  entropy sum relations of Myers-Perry-AdS black holes, and following naturally its new  entropy product relations are given in section four. The last section devotes conclusions and discussions.

\section{Application of the First Law of Thermodynamics And Mathematical Lemma}
\subsection{The First Law}
In this section, by employing the first law of thermodynamics for black hole horizons, we propose two theorems related to the criterion about the angular momentum-independence of entropy sum and product on all horizons. We index which horizon by subscript $i$  and  index which angular momentum by subscript $j$ all through this paper for convenience.
\begin{theorem}\label{1}
For an axisymmetric and stationary rotating black hole with $n\geq2$ horizons and possess $m\geq1$ rotation parameters $a_j$, the entropy sum of all horizons is independent of angular momentum if and only if
\begin{equation}
\frac{\partial(S_1+S_2+S_3+\ldots+S_n)}{\partial J_j}=\frac{\partial{\widetilde S}}{\partial J_j}=-\sum_{i=1}^{i=n}\frac{\Omega_{ij}}{T_i}=0,(j=1,2,\ldots,m)
\end{equation}
\end{theorem}
We can prove this theorem briefly as follows. For a stationary rotating multi-horizons black hole with mass $M$ , electric charge $Q$ and angular momentum $J_j$ in the spacetime background with a cosmological constant $\Lambda$, each horizon possess corresponding Hawking temperature $T_i$, entropy $S_i$, angular velocity $\Omega_{ij}$ related to angular momentum $J_j$, electric potential $\Phi_i$ and thermodynamics volume
$V_i$ related to cosmological constant and we treat it as a dynamical variable \cite{D1,B1,M1,B2,N1,Meng4}. We have the first law of thermodynamics for each horizon, which reads
\begin{equation}
dM=T_idS_i+\Phi_i dQ+\sum_{j=1}^{m}\Omega_{ij}dJ_j+V_id\Lambda
\end{equation}
After transformation, we can get
\begin{equation}
dS_i=\frac{1}{T_i}(dM-\Phi_i dQ-\sum_{j=1}^{m}\Omega_{ij}dJ_j-V_id\Lambda)
\end{equation}
which yields
\begin{equation}
\frac{\partial S_i}{\partial J_j}=-\frac{\Omega_{ij}}{T_i}\label{2}
\end{equation}
so
\begin{equation}
\frac{\partial{\widetilde S}}{\partial J_j}=-\sum_{i=1}^{i=n}\frac{\Omega_{ij}}{T_i},(j=1,2,\ldots,m)
\end{equation}
If entropy sum $\widetilde S$ is angular momentum $J$ independent, it must satisfies the condition that $\frac{\partial \widetilde S}{\partial J_j}=0$ for every $J_j$ related to rotation parameters $a_j$. We have proven theorem (\ref{1}) now.

\begin{theorem}\label{16}
For an axisymmetric and stationary rotating black hole with $n\geq2$ horizons and possess $m\geq1$ rotation parameters $a_j$, the entropy product of all horizons is independent of angular momentum if and only if
\begin{equation}
\sum_{i=1}^{n}\frac{\Omega_{ij}}{T_i S_i}=0,(j=1,2,3,\ldots ,m)
\end{equation}
\end{theorem}
We can easily prove it by using (\ref{2}), and we have
\begin{align}
\frac{\partial \widehat{S}}{\partial J_j}&=\frac{\partial (S_1S_2\ldots S_n)}{\partial J_j}=-\widehat{S}(\frac{\Omega_{1j}}{T_1S_1}+\frac{\Omega_{2j}}{T_2S_2}+\ldots \frac{\Omega_{nj}}{T_nS_n})\\
&=-\widehat{S}\sum_{i=1}^{n}\frac{\Omega_{ij}}{T_iS_i},(j=1,2,3,\ldots ,m)
\end{align}
where $\widehat{S}\neq0$. The proof is finished.

\subsection{The Mathematical Lemma}
The Lemma \cite{Gao1} we are going to introduce is critical to our calculations as we show in the following sections and the general proof is given in \cite{Gao1}.
\begin{lemma}\label{lemma}
Let \{$r_i$\} be n different numbers, then
\begin{equation}
\sum_{i=1}^{n}\frac{r_i^k}{\prod\limits_{j\neq i}^{n}(r_i-r_j)}=0
\end{equation}
where $0\leq k\leq n-2$
\end{lemma}

For example, when $n=3$, it gives
\begin{gather}
\frac{1}{(r_1-r_2)(r_1-r_3)}+\frac{1}{(r_2-r_1)(r_2-r_3)}+\frac{1}{(r_3-r_1)(r_3-r_2)}=0\\
\frac{r_1}{(r_1-r_2)(r_1-r_3)}+\frac{r_2}{(r_2-r_1)(r_2-r_3)}+\frac{r_3}{(r_3-r_1)(r_3-r_2)}=0
\end{gather}
while
\begin{gather}
\frac{r_1^2}{(r_1-r_2)(r_1-r_3)}+\frac{r_2^2}{(r_2-r_1)(r_2-r_3)}+\frac{r_3^2}{(r_3-r_1)(r_3-r_2)}=1
\end{gather}

\section{Entropy Sum of Myers-Perry-AdS Black Holes}
In this section, we investigate the entropy sum of Myers-Perry-AdS black holes by employing theorem (\ref{1}) and lemma (\ref{lemma}). We find that entropy sum of all horizons of Myers-Perry-AdS black holes  is independent of angular momentum of the black holes in all dimensions, including Kerr-Newman-Ads balck holes in four dimensions and BTZ black hole in three dimensions. The angular momentum-independence of entropy sum also holds for Myers-Perry black holes and Kerr-Newman black holes. It is necessary to discuss Myers-Perry-AdS black holes in even dimensions and odd dimensions separately.
\subsection{Even Dimensions}
We suppose that the spactiem dimension $d=2n+2$ with $n\geq1$. The metric form of  Myers-Perry-AdS black holes in Boyer-Linquist coordinates can be expressed as \cite{Dolan1,Pope1}
\begin{equation}
\begin{split}
ds^2&=-W(1+\lambda r^2)dt^2+\frac{2\mu}{U}\left(Wdt-\sum_{i=1}^n\frac{a_i\rho_i^2d\phi}{1-\lambda a_i^2}\right)^2\\
&\quad+\left(\frac{U}{Z-2\mu}\right)dr^2+r^2dy^2+\sum_{i=1}^{n}\left(\frac{r^2+a_i^2}{1-\lambda a_i^2}\right)(d\rho_i^2+\rho_i^2d\phi_i^2)\\
&\quad-\frac{\lambda}{W(1+\lambda r^2)}\left(\sum_{i=1}^{n}\left(\frac{r^2+a_i^2}{1-\lambda a_i^2}\right)\rho_i d\rho_i+r^2ydy\right)^2
\end{split}
\end{equation}
where the functions $W, Z$ and $U$ are
\begin{gather}
W=y^2+\sum_{i=1}^{n}\frac{\rho_i^2}{1-\lambda a_i^2}\\
Z=\frac{(1+\lambda r^2)}{r}\prod\limits_{i=1}^n(r^2+a_i^2)\\
U=\frac{Z}{1+\lambda r^2}\left(1-\sum_{i=1}^{n}\frac{a_i^2\rho_i^2}{r^2+a_i^2}\right)
\end{gather}
The $a_i$ are rotation parameters restricted to $\lambda a_i^2<1$ and $\mu$ is the mass parameter. The parameter $\lambda$ is defined as
\begin{equation}
\lambda=-\frac{2\Lambda}{(d-1)(d-2)}\geq0
\end{equation}
and the rotation parameters $a_i$ are related to angular momentum $J_i$ which appear in the first law via \cite{Dolan1}
\begin{equation}
J_i=\frac{\mu\Omega_{d-2}a_i}{4\pi(1-\lambda a_i^2)\prod_j(1-\lambda a_j^2)}
\end{equation}
where $\Omega_{d-2}$ is the volume of the unit $(d-2)$-sphere
\begin{equation}
\Omega_{d-2}=\frac{2\pi^{\frac{d-1}{2}}}{\Gamma(\frac{d-1}{2})}
\end{equation}

We have horizon function $Z-\mu=0$, which can be rewritten as
\begin{equation}
\prod\limits_{j=1}^{n}(r^2+a_j^2)-2\mu \frac{r}{1+\lambda r^2}=0\label{4}
\end{equation}
and we denote $\prod\limits_{j=1}^{n}(r^2+a_j^2)$ as $\prod(r)$. The horizon function (\ref{4}) has $2n+\epsilon$($n\geq1$) roots and the parameter $\varepsilon$ is defined as
\begin{numcases}{\epsilon=}
 2 & $\lambda\neq0$ \\
 0 & $\lambda=0$
 \end{numcases}

The hawking temperature on the i-th horizon is
\begin{equation}
T_i=\frac{r_i}{2\pi}(1+\lambda r_i^2)\sum_{j=1}^{n}\frac{1}{r_i^2+a_j^2}
+\frac{\lambda r_i^2-1}{4\pi r_i}=\frac{k_i}{2\pi}
\end{equation}
where $r_i$ and $k_i$ denote the corresponding horizon radius and surface gravity, respectively. After calculating, we find that the surface gravity $k_i$ has the following simple and compact expression
\begin{equation}
k_i=\frac{(1+\lambda r^2)^2\partial_r \left(\prod(r)-2\mu\frac{r}{1+\lambda r^2}\right)}{4\mu r}\Big|_{r=r_i}\label{10}
\end{equation}
Angular velocity entering thermodynamical laws corresponding to the j-th rotation parameter on the i-th horizon is
\begin{equation}
\Omega_{ij}=\frac{(1+\lambda r_i^2)a_j}{r_i^2+a_j^2}\label{5}
\end{equation}
We introduce the function
\begin{equation}
f(r)=\frac{(1+\lambda r^2)^2\left(\prod(r)-2\mu\frac{r}{1+\lambda r^2}\right)}{4\mu r}=\frac{(1+\lambda r^2)^2}{4\mu r}\prod\limits_{j=1}^{2n+\epsilon}(r-r_j)\label{3}
\end{equation}
The last term of Eq. (\ref{3}) holds since there exist $2n+\epsilon$ roots of Eq. (\ref{4}). Then we take derivative of function $f(r)$ with respect to $r$
\begin{equation}
\begin{split}
f'(r_i)&=\frac{(1+\lambda r_i^2)^2}{4\mu r_i}\partial_r \left(\prod(r)-2\mu\frac{r}{1+\lambda r^2}\right)\Big|_{r=r_i}=2\pi T_i\\
&=\frac{(1+\lambda r_i^2)^2}{4\mu r_i}\prod\limits_{g\neq i}^{2n+\epsilon}(r_i-r_g)
\end{split}
\end{equation}
which yields
\begin{equation}
\frac{1}{T_i}=\frac{8\pi\mu r_i}{(1+\lambda r_i^2)\prod\limits_{g\neq i}^{2n+\epsilon}(r_i-r_g)}
\end{equation}
Under the consideration of (\ref{5}), we can get
\begin{equation}
\sum_{i=1}^{2n+\epsilon}\frac{\Omega_{ij}}{T_i}=8\pi\mu\sum_{i=1}^{2n+\epsilon}\frac{r_i\frac{a_j}{r_i^2+a_j^2}}{(1+\lambda r_i^2)\prod\limits_{g\neq i}^{2n+\epsilon}(r_i-r_g)}\label{7}
\end{equation}
Note the horizon function (\ref{4}), we have
\begin{equation}
\frac{r_i a_j}{(1+\lambda r_i^2)(r_i^2+a_j^2)}=\frac{a_j}{2\mu}\prod\limits_{k\neq j}^{n}(r_i^2+a_k^2)\label{6}
\end{equation}
Substituting (\ref{6}) into (\ref{7}), then we obtain
\begin{equation}
\sum_{i=1}^{2n+\epsilon}\frac{\Omega_{ij}}{T_i}=4\pi a_j\sum_{i=1}^{2n+\epsilon}\frac{\prod\limits_{k\neq j}^{n}(r_i^2+a_k^2)}{\prod\limits_{g\neq i}^{2n+\epsilon}(r_i-r_g)}=0\label{8},(j=1,2,\ldots,n)
\end{equation}
By using the mathematical lemma (\ref{lemma}), one can easily check that (\ref{8}) holds since the maximal power term of $\prod_{k\neq j}^{n}(r_i^2+a_k^2)$ is $r_i^{2n-2}(2n-2\leq2n+\epsilon-2)$ while the minimal term is $r_i^0$.  We could see that the parameter $\lambda$ related to cosmological constant $\Lambda$ has vanished in (\ref{8}), which means that the cosmological constant may has no effects for the angular-momentum independence of entropy sum. Considering the theorem (\ref{1}), we conclude that the entropy sum of Myers-Perry or Myers-Perry-AdS black holes is angular independent in even dimensions $d\geq4$.

\subsection{Odd Dimensions}
Suppose the spacetime dimension $d=2n+1(n\geq2)$. The metric of Myers-Perry-AdS black holes expressed by  Boyer-Linquist coordinates in the odd dimension is given by
\begin{equation}
\begin{split}
ds^2&=-W(1+\lambda r^2)dt^2+\frac{2\mu}{U}\left(Wdt-\sum_{i=1}^n\frac{a_i\rho_i^2d\phi}{1-\lambda a_i^2}\right)^2\\
&\quad+\left(\frac{U}{Z-2\mu}\right)dr^2+\sum_{i=1}^{n}\left(\frac{r^2+a_i^2}{1-\lambda a_i^2}\right)(d\rho_i^2+\rho_i^2d\phi_i^2)\\
&\quad-\frac{\lambda}{W(1+\lambda r^2)}\left(\sum_{i=1}^{n}\left(\frac{r^2+a_i^2}{1-\lambda a_i^2}\right)\rho_i d\rho_i\right)^2
\end{split}
\end{equation}
where the functions $W, Z$ and $U$ are
\begin{gather}
W=\sum_{i=1}^{n}\frac{\rho_i^2}{1-\lambda a_i^2}\\
Z=\frac{(1+\lambda r^2)}{r^2}\prod\limits_{i=1}^n(r^2+a_i^2)\\
U=\frac{Z}{1+\lambda r^2}\left(1-\sum_{i=1}^{n}\frac{a_i^2\rho_i^2}{r^2+a_i^2}\right)
\end{gather}
The $a_i$ are rotation parameters restricted to $\lambda a_i^2<1$ and $\mu$ is the mass parameter.
The horizon function is
\begin{equation}
\prod\limits_{j=1}^{n}(r^2+a_j^2)-2\mu \frac{r^2}{1+\lambda r^2}=0\label{9}
\end{equation}
The hawking temperature on the i-th horizon is
\begin{equation}
T_i=\frac{r_i}{2\pi}(1+\lambda r_i^2)\sum_{j=1}^{n}\frac{1}{r_i^2+a_j^2}
-\frac{1}{2\pi r_i}=\frac{k_i}{2\pi}
\end{equation}
The surface gravity $k_i$ is
\begin{equation}
k_i=\frac{(1+\lambda r^2)^2\partial_r \left(\prod(r)-2\mu\frac{r^2}{1+\lambda r^2}\right)}{4\mu r^2}\Big|_{r=r_i}\label{17}
\end{equation}
We introduce the function
\begin{equation}
f(r)=\frac{(1+\lambda r^2)^2\left(\prod(r)-2\mu\frac{r^2}{1+\lambda r^2}\right)}{4\mu r^2}=\frac{(1+\lambda r^2)^2}{4\mu r^2}\prod\limits_{j=1}^{2n+\epsilon}(r-r_j)
\end{equation}
After taking derivative of $f(r)$ with respect to $r$, we obtain
\begin{equation}
\begin{split}
f'(r_i)&=\frac{(1+\lambda r_i^2)^2}{4\mu r_i^2}\partial_r \left(\prod(r)-2\mu\frac{r^2}{1+\lambda r^2}\right)\Big|_{r=r_i}=2\pi T_i\\
&=\frac{(1+\lambda r_i^2)^2}{4\mu r_i^2}\prod\limits_{g\neq i}^{2n+\epsilon}(r_i-r_g)
\end{split}
\end{equation}
which yields
\begin{equation}
\sum_{i=1}^{2n+\epsilon}\frac{\Omega_{ij}}{T_i}=8\pi\mu\sum_{i=1}^{2n+\epsilon}\frac{r_i^2\frac{a_j}{r_i^2+a_j^2}}{(1+\lambda r_i^2)\prod\limits_{g\neq i}^{2n+\epsilon}(r_i-r_g)}
\end{equation}
Note the horizon function (\ref{9})
\begin{equation}
\frac{r_i^2 a_j}{(1+\lambda r_i^2)(r_i^2+a_j^2)}=\frac{a_j}{2\mu}\prod\limits_{k\neq j}^{n}(r_i^2+a_k^2)
\end{equation}
Finally we obtain
\begin{equation}
\sum_{i=1}^{2n+\epsilon}\frac{\Omega_{ij}}{T_i}=4\pi a_j\sum_{i=1}^{2n+\epsilon}\frac{\prod\limits_{k\neq j}^{n}(r_i^2+a_k^2)}{\prod\limits_{g\neq i}^{2n+\epsilon}(r_i-r_g)}=0,(j=1,2,\ldots,n)
\end{equation}

The final result is just the same as the case in even dimensions. The entropy sum of BTZ black hole \cite{Meng2} in three dimension is also angular independent. From this point we conclude that the entropy sum of all horizons of rotating black holes is angular momentum independent in all dimensions $d\geq3$

\section{Entropy Product of Myers-Perry-AdS Black Holes}
In this section, we discuss the entropy product of all horizons and we find that the parameter $\lambda$ related to cosmological constant has contribution to the final criterion for the universal property of entropy product while the angular momentum-independence(or dependence) of entropy sum will not be affected by cosmological constant.
\subsection{Even Dimensions}
The area of the i-th horizon is denoted as $A_i$ in even dimension($d=2n+2,n\geq1$)
\begin{equation}
A_i=\Omega_{d-2}\prod\limits_{j=1}^{n}\frac{r_i^2+a_j^2}{1-\lambda a_j^2}
\end{equation}

In Einstein gravity, the Hawking-Bekenstein entropy on the i-th horizon is
\begin{equation}
S_i=\frac{A_i}{4}=\frac{\Omega_{d-2}}{4}\prod\limits_{j=1}^{n}\frac{r_i^2+a_j^2}{1-\lambda a_j^2}\label{11}
\end{equation}

We introduce function $f(r)$
\begin{equation}
\begin{split}
f(r)&=\frac{(1+\lambda r^2)^2\left(\prod(r)-2\mu\frac{r}{1+\lambda r^2}\right)}{4\mu r}\prod\limits_{j=1}^{n}\frac{r_i^2+a_j^2}{1-\lambda a_j^2}\\
&=\frac{(1+\lambda r^2)^2}{4\mu r}\prod\limits_{j=1}^{2n+\epsilon}(r-r_j)\prod\limits_{j=1}^{n}\frac{r_i^2+a_j^2}{1-\lambda a_j^2}
\end{split}
\end{equation}
After taking derivative of $f(r)$ with respect to $r$, together with Eq. (\ref{10}), ( \ref{11}) and ( \ref{4}), we obtain
\begin{equation}
\begin{split}
f'(r_i)&=\frac{(1+\lambda r_i^2)^2}{4\mu r_i}\partial_r \left(\prod(r)-2\mu\frac{r}{1+\lambda r^2}\right)\Big|_{r=r_i}\prod\limits_{j=1}^{n}\frac{r_i^2+a_j^2}{1-\lambda a_j^2}=\frac{8\pi}{\Omega_{d-2}} S_iT_i\\
&=\frac{(1+\lambda r_i^2)^2}{4\mu r_i}\prod\limits_{g\neq i}^{2n+\epsilon}(r_i-r_g)\prod\limits_{j=1}^{n}\frac{r_i^2+a_j^2}{1-\lambda a_j^2}\\
&=\frac{1}{2}\prod\limits_{g\neq i}^{2n+\epsilon}(r_i-r_g)\frac{1+\lambda r_i^2}{\prod\limits_{j=1}^{n}(1-\lambda a_j^2)}\label{12}
\end{split}
\end{equation}
Employing angular velocity formula (\ref{5}), together with (\ref{12}) we can get
\begin{equation}
\begin{split}
\sum_{i=1}^{2n+\epsilon}\frac{\Omega_{ij}}{T_iS_i}=\frac{16\pi}{\Omega_{d-2}}\prod\limits_{f=1}^{n}(1-\lambda a_f^2)
\sum_{i=1}^{2n+\epsilon}\frac{\frac{a_j}{r_i^2+a_j^2}}{\prod\limits_{g\neq i}^{2n+\epsilon}(r_i-r_g)}\label{14}
\end{split}
\end{equation}
From horizon function (\ref{4}), we have
\begin{equation}
\frac{a_j}{r_i^2+a_j^2}=\frac{(1+\lambda r_i^2)a_j}{2\mu r_i}\prod\limits_{k\neq j}^{n}(r_i^2+a_k^2)\label{13}
\end{equation}
Substituting (\ref{13}) into (\ref{14}), then
\begin{equation}
\sum_{i=1}^{2n+\epsilon}\frac{\Omega_{ij}}{T_iS_i}=\frac{8\pi a_j}{\mu\Omega_{d-2}}\prod\limits_{f=1}^{n}(1-\lambda a_f^2)\sum_{i=1}^{2n+\epsilon}\frac{\prod\limits_{k\neq j}^{n}(r_i^2+a_k^2)}{\prod\limits_{g\neq i}^{2n+\epsilon}(r_i-r_g)}\frac{1+\lambda r_i^2}{r_i}\label{15}
\end{equation}
where $\prod\limits_{f=1}^{n}(1-\lambda a_f^2)\neq 0$

We could see that the parameter $\lambda$ does not  vanish which means that we have to discuss (\ref{15}) in two different cases.
\begin{itemize}
\item Case \uppercase\expandafter{\romannumeral1}: $\lambda=0$\\
In this case, there are $2n$ horizons located at $r_i$ which are the roots of horizon function (\ref{4}) and Eq. (\ref{15}) becomes
\begin{equation}
\sum_{i=1}^{2n}\frac{\Omega_{ij}}{T_iS_i}=\frac{8\pi a_j}{\mu\Omega_{d-2}}\sum_{i=1}^{2n}\frac{\prod\limits_{k\neq j}^{n}(r_i^2+a_k^2)}{\prod\limits_{g\neq i}^{2n}(r_i-r_g)}\frac{1}{r_i}
\end{equation}
It's obviously to see that the maximal power term of $r_i^{-1}\prod_{k\neq j}^{n}(r_i^2+a_k^2)$ is $r_i^{2n-3}$ and the minimal is $r_i^{-1}\prod_{k\neq j}^{n}a_k^2$. By employing lemma (\ref{lemma}) and theorem (\ref{16}), the entropy product of rotating black hole is independent of angular momentum if and only if $2n-3\geq0$ and $\prod_{k\neq j}^{n}a_k^2=0$ in even dimensions which means that $n\geq 2$, dimension $d\geq 6$ and at least one and up to $n-1$ rotation  parameter $a_k=0$ since the object we study is a rotating black hole we can't let all rotation parameter be zero.
\end{itemize}

\begin{itemize}
\item Case \uppercase\expandafter{\romannumeral2}: $\lambda>0$\\
In this case, there are $2n+2$ horizons and Eq. (\ref{15}) becomes
\begin{equation}
\sum_{i=1}^{2n+2}\frac{\Omega_{ij}}{T_iS_i}=\frac{8\pi a_j}{\mu\Omega_{d-2}}\prod\limits_{f=1}^{n}(1-\lambda a_f^2)\sum_{i=1}^{2n+2}\frac{\prod\limits_{k\neq j}^{n}(r_i^2+a_k^2)}{\prod\limits_{g\neq i}^{2n+2}(r_i-r_g)}\frac{1+\lambda r_i^2}{r_i}
\end{equation}
It's obviously to see that the maximal power term of $r_i^{-1}(1+\lambda r_i^2)\prod_{k\neq j}^{n}(r_i^2+a_k^2)$ is $\lambda r_i^{2n-1}$ and the minimal is $r_i^{-1}\prod_{k\neq j}^{n}a_k^2$. By employing lemma (\ref{lemma}) and theorem (\ref{16}), the entropy product of rotating black hole is independent of angular momentum if and only if $2n-1\geq0$ and $\prod_{k\neq j}^{n}a_k^2=0$ in even dimensions which means that $n\geq 1$, dimension $d\geq 4$ and at least one and up to $n-1$ rotation  parameter $a_k=0$. While when $n=1, d=4$, there is only one rotation parameter which can't be zero for rotating black hole we study so the entropy product must be angular momentum dependent in four dimensions. We can conclude that, for even dimensions, the entropy product of rotating black holes is angular momentum independent if and only if they exist in dimensions  $d\geq 6$ and possess at leat one zero rotation parameter.
\end{itemize}

\subsection{Odd Dimensions}
The i-th horizon area $A_i$ in odd dimension($d=2n+1,n\geq 2$) has the form
\begin{equation}
A_i=\frac{\Omega_{d-2}}{r_i}\prod\limits_{j=1}^{n}\frac{r_i^2+a_j^2}{1-\lambda a_j^2}
\end{equation}
Hawking-Bekenstein entropy on the i-th horizon is
\begin{equation}
S_i=\frac{A_i}{4}=\frac{\Omega_{d-2}}{4r_i}\prod\limits_{j=1}^{n}\frac{r_i^2+a_j^2}{1-\lambda a_j^2}\label{18}
\end{equation}
We introduce function $f(r)$
\begin{equation}
\begin{split}
f(r)&=\frac{(1+\lambda r^2)^2\left(\prod(r)-2\mu\frac{r^2}{1+\lambda r^2}\right)}{4\mu r^3}\prod\limits_{j=1}^{n}\frac{r_i^2+a_j^2}{1-\lambda a_j^2}\\
&=\frac{(1+\lambda r^2)^2}{4\mu r^3}\prod\limits_{j=1}^{2n+\epsilon}(r-r_j)\prod\limits_{j=1}^{n}\frac{r_i^2+a_j^2}{1-\lambda a_j^2}
\end{split}
\end{equation}
Taking derivative of $f(r)$ with respect to $r$, together with Eq. (\ref{9}), (\ref{17}) and (\ref{18}), it's not hard to get
\begin{equation}
\begin{split}
f'(r_i)&=\frac{(1+\lambda r_i^2)^2}{4\mu r_i^2}\partial_r \left(\prod(r)-2\mu\frac{r}{1+\lambda r^2}\right)\Big|_{r=r_i}\frac{1}{r_i}\prod\limits_{j=1}^{n}\frac{r_i^2+a_j^2}{1-\lambda a_j^2}=\frac{8\pi}{\Omega_{d-2}} S_iT_i\\
&=\frac{(1+\lambda r_i^2)^2}{4\mu r_i^3}\prod\limits_{g\neq i}^{2n+\epsilon}(r_i-r_g)\prod\limits_{j=1}^{n}\frac{r_i^2+a_j^2}{1-\lambda a_j^2}\\
&=\frac{1}{2r_i}\prod\limits_{g\neq i}^{2n+\epsilon}(r_i-r_g)\frac{1+\lambda r_i^2}{\prod\limits_{j=1}^{n}(1-\lambda a_j^2)}\label{19}
\end{split}
\end{equation}
By using angular velocity formula (\ref{5}), horizon function (\ref{9}) and (\ref{19}), then
\begin{equation}
\begin{split}
\sum_{i=1}^{2n+\epsilon}\frac{\Omega_{ij}}{T_iS_i}&=\frac{16\pi}{\Omega_{d-2}}\prod\limits_{f=1}^{n}(1-\lambda a_f^2)
\sum_{i=1}^{2n+\epsilon}\frac{r_i\frac{a_j}{r_i^2+a_j^2}}{\prod\limits_{g\neq i}^{2n+\epsilon}(r_i-r_g)}\\
&=\frac{8\pi a_j}{\mu\Omega_{d-2}}\prod\limits_{f=1}^{n}(1-\lambda a_f^2)\sum_{i=1}^{2n+\epsilon}\frac{\prod\limits_{k\neq j}^{n}(r_i^2+a_k^2)}{\prod\limits_{g\neq i}^{2n+\epsilon}(r_i-r_g)}\frac{1+\lambda r_i^2}{r_i}\label{20}
\end{split}
\end{equation}

The final result we get is just the same as the case in even dimension and we must discuss in two cases, too.
\begin{itemize}
\item Case \uppercase\expandafter{\romannumeral1}: $\lambda=0$\\
In this case, there are $2n$ horizons located at $r_i$ which are the roots of horizon function (\ref{9}) and Eq. (\ref{20}) becomes
\begin{equation}
\sum_{i=1}^{2n}\frac{\Omega_{ij}}{T_iS_i}=\frac{8\pi a_j}{\mu\Omega_{d-2}}\sum_{i=1}^{2n}\frac{\prod\limits_{k\neq j}^{n}(r_i^2+a_k^2)}{\prod\limits_{g\neq i}^{2n}(r_i-r_g)}\frac{1}{r_i}
\end{equation}
It's obviously to see that the maximal power term of $r_i^{-1}\prod_{k\neq j}^{n}(r_i^2+a_k^2)$ is $r_i^{2n-3}$ and the minimal is $r_i^{-1}\prod_{k\neq j}^{n}a_k^2$. By employing lemma (\ref{lemma}) and theorem (\ref{16}), the entropy product of rotating black hole is independent of angular momentum if and only if $2n-3\geq0$ and $\prod_{k\neq j}^{n}a_k^2=0$ in odd dimensions which means that $n\geq 2$, dimension $d\geq 5$ and at least one and up to $n-1$ rotation  parameter $a_k=0$.
\end{itemize}

\begin{itemize}
\item Case \uppercase\expandafter{\romannumeral2}: $\lambda>0$\\
In this case, there are $2n+2$ horizons and Eq. (\ref{20}) becomes
\begin{equation}
\sum_{i=1}^{2n+2}\frac{\Omega_{ij}}{T_iS_i}=\frac{8\pi a_j}{\mu\Omega_{d-2}}\prod\limits_{f=1}^{n}(1-\lambda a_f^2)\sum_{i=1}^{2n+2}\frac{\prod\limits_{k\neq j}^{n}(r_i^2+a_k^2)}{\prod\limits_{g\neq i}^{2n+2}(r_i-r_g)}\frac{1+\lambda r_i^2}{r_i}
\end{equation}
It's obviously to see that the maximal power term of $r_i^{-1}(1+\lambda r_i^2)\prod_{k\neq j}^{n}(r_i^2+a_k^2)$ is $\lambda r_i^{2n-1}$ and the minimal is $r_i^{-1}\prod_{k\neq j}^{n}a_k^2$. By employing lemma (\ref{lemma}) and theorem (\ref{16}), the entropy product of rotating black hole is independent of angular momentum if and only if $2n-1\geq0$ and $\prod_{k\neq j}^{n}a_k^2=0$ in odd dimensions which means that $n\geq 2$, dimension $d\geq 5$ and at least one and up to $n-1$ rotation  parameter $a_k=0$. What's more, for three dimension BTZ black hole, the entropy product can be found in (\cite{Meng2}) which is angular momentum dependent. We can conclude that, for odd dimension, the entropy product of rotating black holes is angular momentum independent if and only if they exist in dimensions $d\geq5$ and posses at least one zero rotation parameter.
\end{itemize}

Together with the discussions for even dimensions and odd dimensions, we obtain a simple conclusion which can be stated as the entropy product of a rotating black holes is angular momentum independent if and only if the black holes exist in dimensions ($d\geq5$) with at leat one zero rotation parameter.

\section{Conclusions and Discussions}
In this paper, we investigate the angular momentum independence of the entropy sum and product for rotating black holes based on the first law of thermodynamics and a mathematical lemma related to Vandermonde determinant. The advantage of the method  is that the explicit forms of the spacetime metric, black hole mass and charge are not needed but the Hawking temperature and entropy formula on the horizons are necessary for static black holes, while our calculations require the expressions of metric and angular velocity formula. We find that the entropy sum is always independent of angular momentum for all dimensions and the angular momentum-independence of entropy product only holds for the dimensions $d>4$ case with at least one rotation parameters $a_i=0$, while the mass-independence of entropy sum and product for rotating black holes only hold for higher dimensions cases ($d>4$) and for all dimensions, respectively. On the other hand, we find that the introduction of a negative cosmological constant does not affect the angular momentum-independence of entropy sum and product but the criterion for angular momentum-independence of entropy product will be affected mathematically. From this point, the entropy sum may be more general than entropy product relations.

We have also obtained some new angular momentum independent thermodynamic relations which are not listed in this paper, though their implying meaning is still waiting for further investigations. We believe with more information
available or coming soon the deep meaning for black hole entropy and its application to some possible quantum gravity effects as well as other aspects for black hole physics related  will be elaborated more clearly, with the hope that we may get some hints for the possible quantum gravity framework to be established.

\section*{Acknowledgements}
For the present work we thank Xu Wei, Wang Deng and Yao Yanhong for helpful discussions. This project is partially supported by NSFC

\bibliography{Angular}
\end{document}